\documentclass{article}

\usepackage{arxiv}

\usepackage[utf8]{inputenc} % allow utf-8 input
\usepackage[T1]{fontenc}    % use 8-bit T1 fonts
\usepackage{hyperref}       % hyperlinks
\usepackage{url}            % simple URL typesetting
\usepackage{booktabs}       % professional-quality tables
\usepackage{amsfonts}       % blackboard math symbols
\usepackage{nicefrac}       % compact symbols for 1/2, etc.
\usepackage{microtype}      % microtypography
\usepackage{lipsum}

\usepackage{appendix}
\usepackage{amsmath}
\usepackage{graphicx}
\usepackage{lineno}
\usepackage{array}
\usepackage{longtable}
\usepackage{algorithm}
\usepackage{algpseudocode}
\usepackage{amssymb}
\usepackage[absolute,overlay]{textpos}

\title{Active-Subspace Analysis of Exceedance Probability for Shallow-Water Waves}

\author{
  Kenan \v Sehi\'c\thanks{Corresponding author: kense@dtu.dk} \\
  Department of Applied Mathematics and Computer Science\\
  Technical University of Denmark\\
  DK-2800 Kgs. Lyngby, Denmark\\
   \And
 Henrik Bredmose\\
  Department of Wind Energy\\
  Technical University of Denmark\\
  DK-2800 Kgs. Lyngby, Denmark\\
  \And
 John D. S\o rensen\\
  Department of Civil Engineering \& Department of Wind Energy\\
  Aalborg University \& Technical University of Denmark\\
  Denmark\\
  \And
 Mirza Karamehmedovi\'c\\
  Department of Applied Mathematics and Computer Science\\
  Technical University of Denmark\\
  DK-2800 Kgs. Lyngby, Denmark\\
}

\begin{document}
\maketitle

\begin{abstract}
We model shallow-water waves using a one-dimensional Korteweg-de Vries equation with the wave generation parameterized by random wave amplitudes for a predefined sea state. These wave amplitudes define the high-dimensional stochastic input vector for which we estimate the short-term wave crest exceedance probability at a reference point. For this high-dimensional and complex problem, most reliability methods fail, while Monte Carlo methods become impractical due to the slow convergence rate. Therefore, first within offshore applications, we employ the dimensionality reduction method called \textit{Active-Subspace Analysis}. This method identifies a low-dimensional subspace of the input space that is most significant to the input-output variability. We exploit this to efficiently train a Gaussian process that models the maximum 10-minute crest elevation at the reference point, and to thereby efficiently estimate the short-term wave crest exceedance probability. The active low-dimensional subspace for the Korteweg-de Vries model also exposes the expected incident wave groups associated with extreme waves and loads. Our results show the advantages and the effectiveness of the active-subspace analysis against the Monte Carlo implementation for offshore applications.
\end{abstract}

% keywords can be removed

\section{Introduction}
\label{intro}
Nonlinear hydrodynamic effects are a major concern in bottom-fixed and floating offshore structures at shallow and intermediate depth. Structures such as wind turbines must be designed to withstand extreme nonlinear waves with strongly nonlinear behavior. The simplest model of the waves would stem from linear wave theory and use a Gaussian stochastic model for the wave surface, resulting in a Gaussian response. However, this approach ignores the marked asymmetry in the waves, which means that the wave crest elevation systematically exceeds the trough depths at the same probability level \cite{win3}. The asymmetry increases with decreasing water depth, which eventually produces substantial instabilities resulting in breaking waves and extreme loads. A number of uncertainty sources need to be accounted for when applying numerical wave simulations as an attempt to represent the real offshore conditions \cite{bigoni,samocita,ge,samocita2,samocita3,turk}. These uncertainties are related to the long-term representation of sea-state parameters, wave surface elevation, kinematics, and estimation of wave loads.

For structural reliability analysis, the probability of failure, in general, is written as a $d$-fold integral
\begin{equation}\label{pf}
    P_F = \int_{g(\theta)\leq 0} \pi_d(\theta) d\theta,
\end{equation}
where $\theta \in \mathbb{R}^{d}$ is the uncertain input of a numerical model for the limit-state function $g(\theta)$, $\pi_d$ is the joint probability density function (PDF) for $\theta$, and $g(\theta)\leq 0$ is the failure criterion.

For failure modes within the offshore engineering framework, $g(\theta)$ can model failure events related to wave load effects exceeding arbitrary specified resistances. We here assume the failure event to be related to the maximum crest elevation exceeding a critical level within a certain sea state. We choose the sea-state duration of 10 minutes, which can be relevant for offshore wind turbines. Eq.~\eqref{pf} is related to the short-term exceedance probability as standard normal random variables $\theta$ construct random wave amplitudes for the wave generation with a predefined ocean-wave spectrum and the wave propagation time. If we would additionally include uncertainties/variability related to the sea state, we would evaluate the long-term exceedance probability $P_L$ related to, e.g., one year as
\begin{equation}
    P_L = \int_{\text{state}} P_F(\text{state}) \pi(\text{state}),
\end{equation}
where $P_F(\text{state})$ is the probability of failure for a given sea state obtained by Eq.~\eqref{pf} and $\pi(\text{state})$ accounts for the long-term stochastic modeling of the sea state parameters. In Eq.~\eqref{pf}, we assume $\theta$ is a standard normal variable. If this is not the case, the Rosenblatt transformation \cite{rosen} or the Nataf distribution \cite{nataf} can be used to transfer a non-standard input distribution to the standard normal space. In our study, we focus on the short-term exceedance probability $P_F$, Eq.~\eqref{pf}, for a predefined sea state with independent and identically distributed (iid) random variables $\theta$ drawn from the standard normal density $\pi_d$. We model the wave surface elevation but do not include the effects of model uncertainties in the estimations. Further, we formulate the limit-state function such that failure corresponds to the 10-minute maximum crest elevation exceeding a threshold value $\gamma$, and the failure condition is rewritten as $g(\theta)\geq\gamma$.

The standard reliability approach based on FORM/SORM fails for multiple design points and high-dimensional cases \cite{form}. A more robust approach would be to use the simple Monte Carlo (MC) method that can handle any numerical model. The simple MC approximates Eq.~\eqref{pf} by the sample mean of the indicator function $\mathbb{I}(\theta)$, where $\mathbb{I}(\theta) = 1$ if $g(\theta)\geq \gamma$ and $\mathbb{I}(\theta) = 0$ otherwise. The major disadvantage of MC is its inefficiency. Following the mean squared error indicator for a finite sampling of Eq.~\eqref{pf} \cite{mcbook}, we would need to evaluate a numerical model $5\cdot10^4$ times to estimate the exceedance probability of $2\cdot 10^{-3}$ with the relative error less than $0.1$. It would take approximately 35 days to estimate the sample mean of Eq.~(\ref{pf}) for a numerical model that runs for 1 minute. Specific variance reduction and surrogate approximation methods such as Polynomial Chaos expansion \cite{xiuli,turk} and Gaussian (Kriging) process \cite{bruno} were proposed to improve the performance. However, their requirements would exponentially grow with the dimension. For Gaussian process regression, a large covariance matrix would need to be inverted several times to produce a prediction.

Therefore, a solution is to search for and exploit a low-dimensional subspace of the input space of initial uncertainties that captures the variability of the limit-state function and that constitutes a suitable low-dimensional foundation for surrogate models. This method is called active-subspace analysis (ASA)~\cite{paul1}. Previously, similar work had been done in the Ph.D. thesis by Trent M. Russi \cite{russi}. It is based on the gradients of the system output, in our case the gradients of the limit-state function, and it can be seen as a principal component analysis in the input space. The gradients can reveal hidden correlation between linear combinations of the input parameters $\theta$ of a numerical wave model $g(\theta)$ and the variability of the quantity of interest, e.g., the maximum crest elevation. We hence determine a low-dimensional subspace by rotating the input space, separating the directions of substantial variability from directions where the limit-state function changes insignificantly on average \cite{paul2}. Gradients can be estimated numerically by adjoint methods \cite{brian,adjoint1}, finite difference approximations or automatic differentiation \cite{brian,ad1}. For this paper, we employ forward automatic differentiation (F-AD). In high-dimensional numerical experiments, F-AD is inefficient as it requires one realization per input parameter. However, combining F-AD with an adjoint equation, gradients for all input parameters can be estimated within one numerical realization. We here do not include the adjoint approach. We apply the active-subspaces method on a simplistic, yet nonlinear, shallow-water wave model that is a reasonable intermediate step toward a fully nonlinear model. This model is thus used here to test the advantages and disadvantages of the active-subspace analysis within offshore applications against the standard methods.

We examine the implementation of the active-subspace analysis within Gaussian process regression to efficiently and accurately evaluate the short-term exceedance probability for the maximum 10-minute crest elevation at the reference point. Section \ref{section:GP} briefly introduces Gaussian process regression, while Section \ref{section:ASA} outlines the theoretical background of active-subspace analysis. In Section~\ref{section:kdv22}, we describe the shallow-water wave model, and Section~\ref{section:results} contains the numerical results. There, we demonstrate that Gaussian process regression based on the active-subspace analysis can estimate the exceedance probability based on only $1\%$ of the required Monte Carlo evaluations. The paper closes with the conclusion in Section~\ref{section:end}.

\section{Gaussian process regression}\label{section:GP}
Expensive numerical models are often evaluated at only a few carefully designed points, and the results are then used to formulate cheap surrogate models. Gaussian process regression (GP, also known as kriging) is a standard surrogate approach that improves the performance of simple and efficient polynomial regressions by including a probability distribution over the sample set with a kernel function. In general, we estimate the most probable form of a function based on the training data and simple polynomial regression. This typically does not require a large sample set, even for high-dimensional problems. The uncertainty measure for predictions is found from the confidence interval, which is very useful for the sequential design and generally as an error indicator.

Gaussian process regression describes a smooth function $g(\theta)$ as a realization of an underlying Gaussian process \cite{bruno}

\begin{equation}
    g(\theta) \approx \hat{g}(\theta) = \beta^T \cdot f_T(\theta) + \sigma^2 Z_{GP}(\theta,\omega_z),
\end{equation}
where $\beta^T \cdot f_T(\theta)$ is the trend of the GP which is a simple regression form, e.g., linear or quadratic, $\sigma^2$ is the Gaussian process variance and $Z_{GP}(\theta,\omega_z)$ is a zero-mean, unit-variance stationary Gaussian process with $\omega_z$ an elementary event in the probability space. The trend describes the global behavior of a function $g(\theta)$. The probabilistic foundation of a Gaussian process is a kernel matrix $\mathbf{K}_{ij} = K(|\theta_i - \theta_j|;\Theta)$ with hyperparameters $\Theta$ (such as the overall correlation of samples or smoothness). The overall performance is sensitive to the selection of the optimal kernel function and of the design points. Generally, finding an optimal number of design points $N$ for Gaussian process regression is a standard challenge. Gramacy and Apley \cite{gramacy} suggested selecting the number of design points which minimizes the mean squared predictive error.

Define the input matrix $\mathbf{X} = (\theta_{ij}) \in \mathbb{R}^{N \times d}$ and write the corresponding evaluations of a numerical model $g(\theta)$ as $Y = (Y_i = g(\theta_i)) \in \mathbb{R}^{N\times 1}$. Firstly, the parameters $\beta, \sigma^2$ are generated by a generalized least-squares regression \cite{bruno}. For a kernel matrix $\mathbf{K}_{ij}$, the hyperparameters $\Theta$ are estimated by the maximum likelihood estimation. Finally, for predictions, we define the prediction mean $\mu_g (\theta)$ and the corresponding variance $\sigma_g^2(\theta)$ for a numerical model $g(\theta)$ as \cite{bruno}

\begin{equation}\label{gpm}
    \mu_g(\theta) = f_T(\theta)\cdot\beta + k(\theta)^T\mathbf{K}^{-1}(Y - \mathbf{F}_T\beta),
\end{equation}

\begin{equation}\label{gps}
    \sigma_g^2(\theta) = \sigma^2\Bigg( 1 - \langle f_T(\theta)^Tk(\theta)^T \rangle \begin{bmatrix}
0 & \mathbf{F}_T^T\\
\mathbf{F}_T & \mathbf{K} \end{bmatrix}^{-1} \begin{bmatrix}
f_T(\theta) \\
k(\theta) \end{bmatrix} \Bigg).
\end{equation}
Here $k(\theta)$ is the correlation between the prediction and the rest of the samples within the set and $\mathbf{F}_T$ is the information matrix regarding the GP trend. Now, instead of using an expensive numerical model $g(\theta)$ to evaluate, e.g., the maximum crest elevation at an offshore application, we can use a cheap surrogate model, Eq.~\eqref{gpm}, and estimate the short-term exceedance probability, Eq. ($\ref{pf}$), by simple MC. The second moment, Eq.~\eqref{gps}, quantifies uncertainties in the predictions. The MATLAB function \texttt{fitrgp} from the Statistics and Machine Learning Toolbox trains a Gaussian process regression model based on design points.

However, for higher dimensions, e.g., $d=100$, the process of estimating the GP parameters becomes time-consuming as it requires repeated inversion of the $N \times N$ kernel matrix, incurring a $\mathcal{O}(N^3)$ cost. Also, to estimate the hyperparameters with the maximum likelihood approach, the kernel matrix $K_{ij}$ needs to be inverted. The process can be improved if we find a low-dimensional optimal representation of $\theta$ for $g(\theta)$. We assume that it is inexpensive to estimate gradients numerically for a numerical model $g(\theta)$.

\section{Active-subspace analysis}\label{section:ASA}
Active-subspace analysis (ASA) is a dimensionality reduction approach that has been studied in detail in the book \cite{paul1} by Paul G. Constantine. It is based on identifying and exploiting the most important linear combinations of the input parameters concerning the quantity of interest, e.g., the maximum crest elevation at the reference point. A split between important and less important linear directions in the input space is usually defined by a spectral gap in the eigenvalues of the gradient data. 

We assume that $g:\mathbb{R}^{d}\rightarrow \mathbb{R}$ is a continuous and differentiable function that is square integrable with respect to a probability density $\pi_d$ for the initial uncertainties $\theta$. An active subspace, i.e., a subspace of the input space with significant variation of the output, is typically spanned by a relatively small number $(\ll d)$ of eigenvectors of the symmetric positive semi-definite $d \times d$ matrix $\mathbf{C}$, which is an uncentered covariance matrix of the output gradients. Thus, we write the expected value of the outer product of the gradients as \cite{paul1,paul2,paul3}
\begin{equation}\label{C}
    \mathbf{C} = \int \nabla_{\theta} g(\theta) \nabla_{\theta} g(\theta)^{T} \pi_d(\theta) \textit{d} \theta = \mathbf{W} \boldsymbol{\Lambda} \mathbf{W}^T,
\end{equation}
where $g(\theta)$ is the quantity of interest, $\nabla_{\theta} g$ is the gradient of $g(\theta)$ with respect to $\theta$, the non-negative eigenvalues of $\textbf{C}$ are sorted in descending order along the diagonal of the diagonal matrix $\boldsymbol{\Lambda}$, and $\mathbf{W}$ is the orthogonal matrix of eigenvectors $d \times d$.

As shown in \textbf{Lemma 3.1.} \cite[p. 23]{paul1}, each eigenvalue $\lambda_{i}$ is the expected squared directional derivative of $g(\theta)$ along the corresponding eigenvector $\mathbf{w}_{i}$,
\begin{equation}\label{lam}
    \lambda_{i}=\int (\nabla_{\theta} g(\theta)^{T} \mathbf{w}_{i})^{2} \pi_d(\theta) \textit{d} \theta.
\end{equation}
Hence, if there is a significant spectral gap after the first largest $r$ eigenvalues of $\textbf{C}$, with $\mathbf{W}_r$ being the first $r$ columns of the orthogonal eigenvector matrix $\mathbf{W}$, then it should be possible to construct a reasonable approximation of $g(\theta)$ in terms of \cite{paul1,paul2}
\begin{equation}\label{g}
    g(\theta)\approx \hat{g}(\mathbf{W}_r^{T} \theta),
\end{equation}
where $\hat{g}$ is a surrogate model obtained using, e.g., a regression. The reduction of the input space dimension helps quantify uncertainties in an otherwise infeasible setting.

\subsection{Active subspace estimation}\label{regress}
The covariance matrix $\mathbf{C}$, Eq.~\eqref{C}, cannot be computed exactly. Therefore, we employ the simple Monte Carlo method to approximate it as \cite{paul1,paul2} 

\begin{equation}\label{CC}
    \mathbf{C} \approx \mathbf{\hat{C}} = \frac{1}{M} \sum^{M}_{i=1} (\nabla_{\theta_{i}} g({\theta}_{i}))(\nabla_{\theta_{i}} g({\theta}_{i}))^T.
\end{equation}

The estimation of how many samples are required to approximate the covariance matrix accurately is heuristic. At least, to have full rank, we need to have $M=d$. Constantine~\cite[p. 35]{paul1} recommends $M=\alpha_A k_A \log(d)$, where $\alpha_A$ is an oversampling factor between 2 and 10, and $k_A$ is the number of eigenvalues to approximate.  If we can evaluate the gradients analytically, it is straightforward to use Eq.~\eqref{CC}. However, this is not the case with numerical models in general. At least, we can approximate the gradients. First-order finite differences (FD) require $M\cdot(d+1)$ model evaluations per gradient evaluation, which is infeasible for high-dimensional computations. Instead, to employ the FD approach, we use forward automatic differentiation, as described in Section \ref{AD}. The active-subspace analysis based on the singular value decomposition is outlined in \textbf{Algorithm \ref{al12}}.

\begin{algorithm}
\caption{Monte Carlo Estimation of Active Subspace~\cite{paul1,paul2}}
\label{al12}
\begin{algorithmic}[1]
\Procedure{ASA}{$g(\theta)$,$\pi_d$}
    \State Draw $M$ iid $\theta_i$ from $\pi_d$. // Use $M=\alpha_A\cdot k_A\cdot \log{(d)}$. To have at least a full matrix rank, we should have $M\ge d$. 
    \State For each $\theta_i$, define $\nabla_{\theta}g_i = \nabla_{\theta}g(\theta_i)$. // Use an adjoint solver or a finite difference approach.
    \State Define the matrix $\mathbf{G}$ following the relation $\hat{\mathbf{C}} = \mathbf{G}\mathbf{G}^T$ as
    \begin{equation}
        \mathbf{G} = \frac{1}{\sqrt{M}} \Bigg[ \nabla_{\theta_1} g(\theta_1), \nabla_{\theta_2} g(\theta_2),...,\nabla_{\theta_N} g(\theta_N)\Bigg].
    \end{equation}
    \State Compute the singular value decomposition $\mathbf{G}= \mathbf{\widehat{W}} \sqrt{\mathbf{\widehat{\boldsymbol{\Lambda}}}} \mathbf{\hat{V}}^{T}$.
\EndProcedure
\end{algorithmic}
\end{algorithm}

Following \textbf{Line 5} in \textbf{Algorithm \ref{al12}}, we search for a spectral gap in the singular values of the matrix $\mathbf{G}$ as a means of identifying the important (active) and the unimportant (inactive) directions in the input space $\theta$. If the singular values do not present a significant spectral gap, an alternative is to use Eq.~\eqref{dist} to estimate the distance between the true $r$-dimensional active subspace and the estimated $r$-dimensional active subspace. This estimation also depends on the spectral gap $\lambda_r-\lambda_{r+1}$. The larger this gap is, the better the estimate, since \cite[p. 32]{paul1}

\begin{equation}\label{dist}
    \text{dist}(\text{ran}(\mathbf{W}_r),\text{ran}(\widehat{\mathbf{W}}_r) \leq \frac{4\lambda_1\mathbf{\epsilon}}{\lambda_r-\lambda_{r+1}},
\end{equation}
where $\mathbf{W}_r$ is the true subspace, $\mathbf{\widehat{W}}_r$ is the estimated subspace, the denominator is the spectral gap, 'ran' with a matrix argument is a shorthand notation for the range of the columns of the matrix, and $\mathbf{\epsilon}$ is the relative accuracy. The relative accuracy can be estimated, following \textbf{Corollary 3.10} \cite[p. 32]{paul1}, by
\begin{equation}
        \mathbf{\epsilon} \leq \frac{\lambda_r-\lambda_{r+1}}{5\lambda_1}.
\end{equation}
As pointed out by Constantine \cite[p. 32]{paul1}, the bound in \textbf{Corollary 3.10} could perhaps be improved. Nevertheless, we can use this estimate to bound the error in the estimated subspaces. The subspace approximation error is inversely proportional to the corresponding gaps in the singular values. Therefore, for example, the estimate of a three-dimensional active subspace is more accurate than the estimate of a two-dimensional active subspace, if the spectral gap is larger between $\lambda_3$ and $\lambda_4$ than between $\lambda_2$ and $\lambda_3$.

\subsubsection{Gradient approximations by Automatic Differentiation (AD)}\label{AD}
To construct the gradients for the active-subspace analysis without using the FD approach, forward automatic differentiation (AD) is applied on subroutine by subroutine basis to the code required to compute the quantity of interest. The main strategy behind AD is to define the input parameter $\theta$ with an additional second component, $\theta + \dot{\theta} \Gamma$. Here, $\Gamma$ is a symbol distinguishing the second component analogous to the imaginary unit $i=\sqrt{-1}$, but in the AD case $\Gamma^2=0$ as opposed to $i^2=-1$. The input parameters have been converted from type "real" to type "complex". The "real" part will remain unchanged, and the "imaginary" part can be used to approximate the derivative of variables for a single design variable. We add an imaginary perturbation to the desired complex input parameters to determine the corresponding imaginary part of the quantity of interest. When the process is generated and validated, forward differentiation can be performed. In this paper, the forward differentiation is done using the ADiMat software from the Institute for Scientific Computing of TU Darmstadt \cite{adimat}.

\subsubsection{Constructing a regression surface} 
Once the spectral gap is identified, the limit-state function $g(\theta)$ is replaced by its low-dimensional surrogate by expressing the initial uncertainties ${\theta}\in \mathbb{R}^d$ in terms of the active part $\{y_A\}$ and inactive part $\{z_A\}$, \cite[p. 24]{paul1}
\begin{equation}
    {\theta}=\mathbf{W}\mathbf{W}^{T}\theta=\mathbf{W}_{r}\mathbf{W}^{T}_r\theta+\mathbf{W}_{d-r}\mathbf{W}^{T}_{d-r}\theta= \mathbf{W}_ry_A+\mathbf{W}_{d-r}z_A.
\end{equation}
In particular, this means that $g(\theta)$ is expressed $g(\mathbf{W}_{n}y_A+\mathbf{W}_{m-n}z_A)$. Small perturbation of $z_A$ changes $g(\theta)$ insignificantly on average. Thus, the optimal approximation of $g(\theta)$ is to calculate the conditional expectation for each fixed $y_A$, and we define $\hat{g}(y_A)$ as
\begin{eqnarray}\label{g(y)}
\hat{g}(y_A)=\int g(\mathbf{W}_ry_A+\mathbf{W}_{d-r}z_A) \pi_{z_A|y_A}(z_A) dz_A,
\end{eqnarray}
where $\pi_{z_A|y_A}(z_A)$ is a conditional probability density~\cite[p. 49]{paul1}. One can argue that we are going back to multidimensional integration again, however using MC has its advantages in this specific case as the variation of $g(\theta)$ in the inactive subspace is significantly small and requires only a small number of samples. Therefore, we write $g(y_A)$ based on MC as 
\begin{equation}
    \hat{g}(y_{A, j}) \approx \frac{1}{Z} \sum^{Z}_{i=1}g(\mathbf{W}_ry_{A, j} + \mathbf{W}_{d-r}z_{A,i}),
\end{equation}
where $Z$ is the number of samples in the inactive directions and $\{z_{A,i}\}$ are random sample points from the conditional probability density $\pi_{z_A|y_A}(z_A)$ \cite[p. 51]{paul1}. If the function $g(\theta)$ is constant in an inactive directions, meaning that the eigenvalue for this direction is zero, then we need to sample only once to account properly for the variation of $g(\theta)$ along this direction.

Hence, to construct a low-dimensional approximation of $g(\theta)$, we generate a number $N_y$ of fixed points $y_{A,j}$ in the active subspace and collect their corresponding conditional expectations $\{\hat{g}(y_{A, j})\}$. Based on the pairs $\{y_{A, j}, \hat{g}(y_{A, j})\}$ along the active directions $\mathbf{W}_r$, we generate a regression surface for $\hat{g}({y_A})$ that is a low-dimensional approximation of the limit-state function $g(\theta)$,
\begin{equation}
    g(\theta)\approx \hat{g}(\mathbf{W}^T_r\theta).
\end{equation}
Thus, instead of training a Gaussian process model in the original, highly dimensional space $\mathbb{R}^d$, we first project the training set onto the active, low-dimensional subspace $\mathbb{R}^r$ ($r \ll d)$ using $\mathbf{W}_r^T$ and then train a Gaussian process model efficiently and accurately between $\mathbf{W}^T_r\theta \in \mathbb{R}^r$ and $Y \in \mathbb{R}$.

\section{A simple 1D Korteweg-de Vries model}\label{section:kdv22}
While our long-term goal is the accelerated load statistics for fully nonlinear models, we here use a much simpler wave model to investigate the feasibility of the active-subspace analysis for rare events.

\begin{figure}[ht]
    \centering
    \includegraphics[scale=0.4]{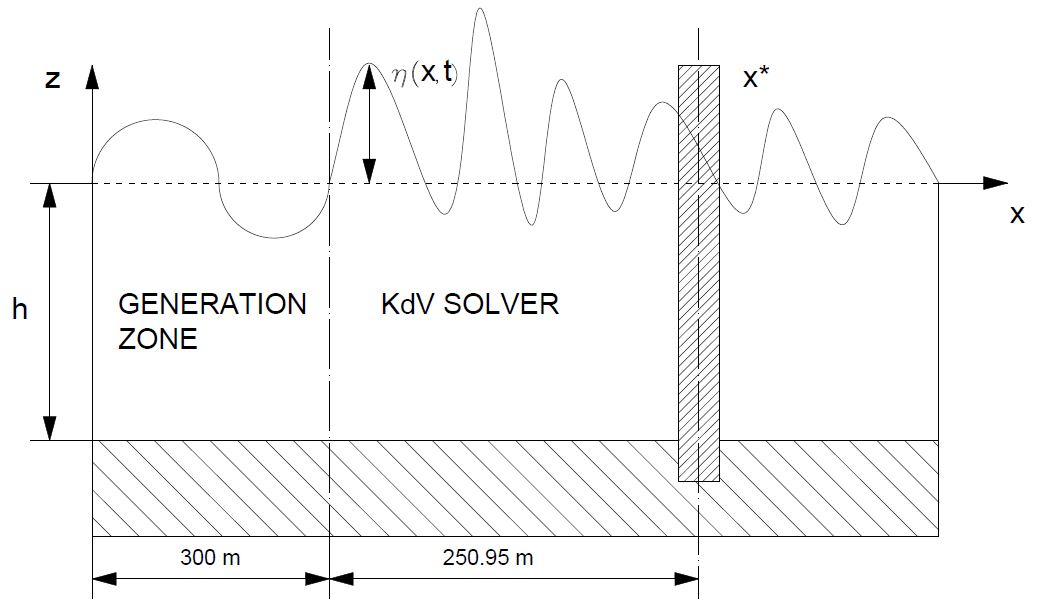}
    \caption{Numerical KdV22 shallow-water waves model for the fixed spatial location $x^*$.}
    \label{kdv_draw}
\end{figure}

We consider unsteady water waves defined by the Korteweg-de Vries equation (KdV) for one-dimensional nonlinear surface flows under the influence of gravity. KdV, derived by Korteweg and de Vries (1895), describes weakly nonlinear shallow-water waves by adding one dispersive term to the nonlinear shallow water equation. There are different modifications of the KdV equation, and we here use KdV22 \cite{henrik},

\begin{equation}\label{KdV22}
%\begin{split}
    \eta_{t}(x,t)+\sqrt{g\cdot h}\cdot\eta_{x}(x,t)+ \frac{3}{2} \sqrt{\frac{g}{h}}\eta(x,t)\eta_{x}(x,t)+(\beta+\frac{1}{6})\sqrt{\frac{g}{h}}h^{3}\eta_{xxx}(x,t)+\beta h^{2}\eta_{xxt}(x,t)=0,
%\end{split}
\end{equation}
with $\beta=19/60$. The linear phase speed for this choice of $\beta$ is the Pad\'e [2,2] approximation of the fully disperse result, $h$ is the seabed depth, $x$ is the spatial-domain variable, $g$ is the gravitational acceleration, $\eta_t$ represents $(\partial\eta/\partial t)(x,t)$ and $\eta_x$ represents $(\partial\eta/\partial x)(x,t)$. The term $\eta_t$ describes the temporal evolution of unidirectional waves, the nonlinear term $\eta\eta_x$ accounts for the steepening of the wave, $\eta_{xxx}$ is a linear dispersive term, and $\eta_{xxt}$ describes the spreading of the waves. For the KdV22 model, which does not describe breaking waves, we assume inviscid and irrotational flow. The seabed is assumed to be flat at the depth of $h=20$m. The sketch of the numerical domain is shown in Fig. \ref{kdv_draw}.

We solve Eq.~\eqref{KdV22} by splitting temporal and spatial derivatives and extending the problem periodically along the $x$-axis. We neglect the spectral content above $60\%$ of the Nyquist frequency to avoid aliasing from the quadratic nonlinearity. For the spatial domain, we employ the classical fourth-order Runge-Kutta method. The generation zone damps the numerical solution $\eta$ that propagates into the zone at its 'outer edge' and transforms it continuously to the desired signal $\eta_{\rm BC}$ out of the zone at its 'inner edge,' by enforcing the correction
\[
\eta:=\eta-\gamma_{\rm force}\chi(\xi)(\eta-\eta_{\rm BC}),
\]
where $\gamma_{\rm force}=3.5$, $\chi$ is the spatial weighting factor~\cite{henrik2}
\begin{equation}
    \chi(\xi)=1-\frac{\exp(\xi^{\beta_{\rm shape}})-1}{\exp(1)-1},
\end{equation}
and $\beta_{\rm shape}=3.5$ is a wave shape factor. Finally, $\xi \in [0,1]$ is a local coordinate, equal to zero at the outer edge and to one at the inner edge of the generation zone.

\begin{figure}[ht]
    \centering
    \includegraphics[scale=0.35]{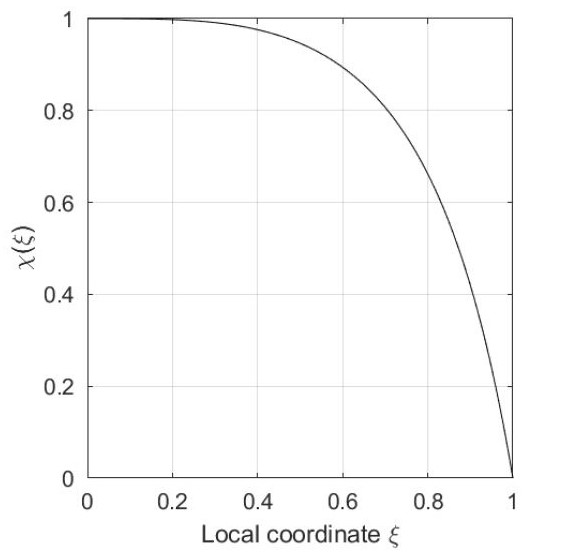}
    \caption{Graph of the spatial weighting factor, $\chi(\xi)$, used for matching the periodic boundary condition \cite{henrik2}.}
    \label{gen}
\end{figure}

\subsection{Wave generation}\label{bc}
Ocean waves are stochastic and can be reasonably well described as Gaussian and ergodic responses. This description provides a good starting point for numerical wave simulations. Therefore, the one-dimensional initial surface elevation used in this paper as the boundary condition is
\begin{equation}\label{KC}
    \eta(x,t)=\sum^{d/2}_{j=1}\sqrt{S(f_j)\cdot \Delta f}\Bigg[A_{j}\cos(\omega_{j}t-k_{j}x)+B_{j}\sin(\omega_{j}t-k_{j}x)\Bigg].
\end{equation}
Here, $S(f_j)$ is the JONSWAP spectrum (Section~\ref{jonswap}), $f_j$ is the frequency, $\Delta f = 1/T$ is the inverse of the wave simulation duration $T$, and $A_j$ and $B_j$ are random variables drawn from the standard normal distribution $\mathcal{N}(0,1)$. For the active-subspace analysis, we define $\theta\in\{ (A_1,\dots,A_{d/2},B_1,\dots,B_{d/2})\} \subseteq \mathbb{R}^d$. The frequency step $\Delta f$ determines the value of $d$. For example, for 1-hour wave propagation, Eq.~\eqref{KC} requires $d \approx 1802$, which results in a highly complex uncertainty quantification problem. 

\subsubsection{Wave spectrum}\label{jonswap}

The wave spectrum density $S(f_j)$ describes the power spectrum of the free surface elevation. There are many wave spectra used for offshore applications in deep water. A fundamental spectrum is the Pierson–Moskowitz spectrum (PM), which describes a fully developed sea. PM is used for fatigue analysis and extreme analysis. We write \cite{standard}

\begin{equation}
    S_{PM}(f_j)=0.3125\cdot H_{S}^{2}\cdot f_P^{4}\cdot f_j^{-5} \cdot \exp\Bigg(-1.25\cdot \Bigg(\frac{f_P}{f_j}\Bigg)^{4}\Bigg),
\end{equation}
where $H_S$ is the significant wave height [m], $f_P$ is the peak frequency [Hz] related to the peak period $T_P$ by $f_P=1 /T_P$ and $f_j$ is the corresponding frequency [Hz].

\begin{figure}[ht]
    \centering
    \includegraphics[scale=0.3]{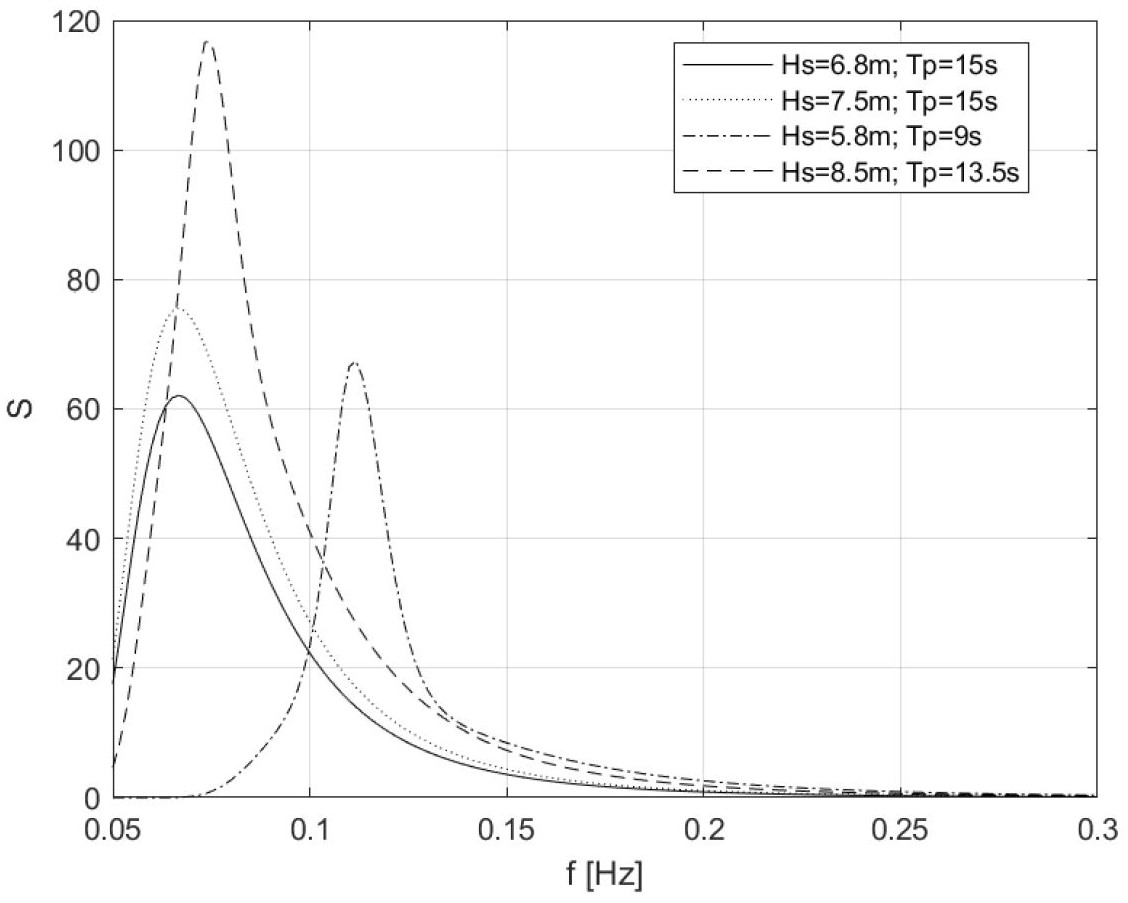}
    \caption{JONSWAP spectrum for different significant wave heights $H_{S}$ and wave periods $T_P$.}
    \label{js}
\end{figure}
The JONSWAP (JS) spectrum is a modification of the PM spectrum for a developing sea state in a fetch limited interaction. JS accounts for a higher peak and a narrower spectrum in a storm situation. Hence, it is often used for extreme events analyses~\cite{turk}. JS has additional two parameters: a peak enhancement factor $\gamma^{\alpha}$ and a normalizing factor $C_{JS}(\gamma)$. Here $\gamma^{\alpha}$ increases the peak and narrows the spectrum, and $C_{JS}(\gamma)$ reduces the spectral density to ensure the energy balance. Thus, we write \cite{standard}

\begin{equation}
    S(f_j)=C_{JS}(\gamma)\cdot S_{PM}(f_{\eta})\cdot \gamma^{\alpha}.
\end{equation}

Figure \ref{js} shows examples of the JS spectrum energy distribution curve with different significant wave heights $H_S$ and time periods $T_P$. We can see that JS is a narrow-banded spectrum. Its energy is mainly focused in a certain frequency band.

\section{Results}\label{section:results}
The KdV22 shallow-water wave model, Eq.~\eqref{KdV22}, is not fully nonlinear but still represents a good intermediate step toward a fully nonlinear model. Expensive numerical wave models such as OceanWave3D \cite{allan} require a substantial computational effort to produce reference results, due to the slow convergence rate of MC methods. Thus, it is natural to use a simple but representative replacement such as KdV22 to test and investigate the active-subspace analysis.

\begin{figure}[ht]
    \centering
    \includegraphics[scale=0.28]{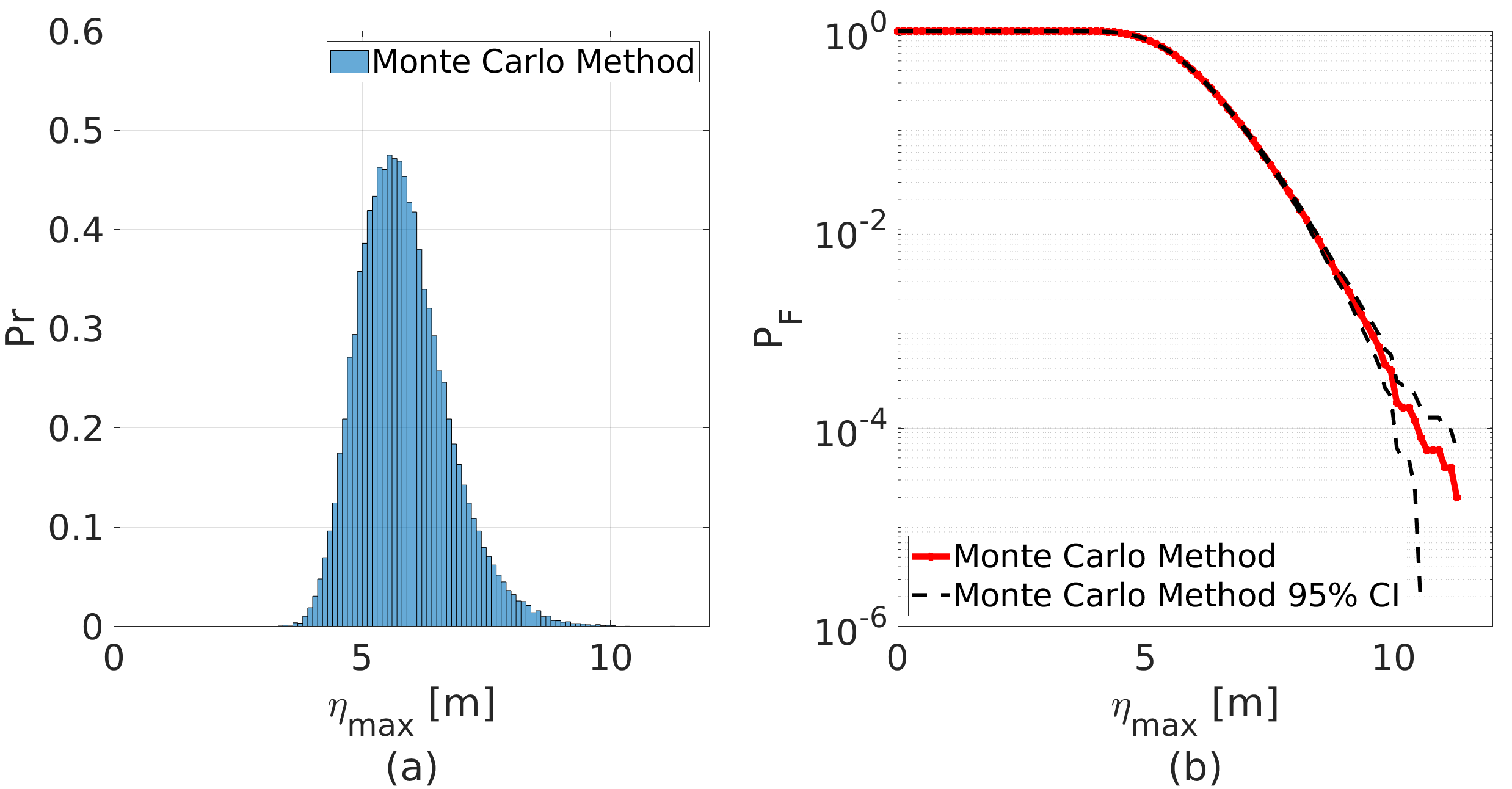}
    \caption{(a) The probability density function for $\eta_{\rm max}$. (b) The short-term exceedance probability of $\eta_{\rm max}$ for 10 minutes, based on $N_{MC} = 5\cdot 10^4$.}
    \label{glavno}
\end{figure}

In our study, unidirectional water waves propagate in a predefined sea state for $T = 600$ seconds. The usual length of a predefined sea state is 1 hour or 3 hours. We use the length of $10$ minutes due to computation limitations. The idea is to have a fast solver to test different approaches before implementing an expensive, fully nonlinear model. Usually, $10$ minutes are used for wind load modeling as a time interval with stationary conditions for the wind field turbulence. The significant wave height and the peak period have been specified as $H_{s}=6.8$ meters and $T_{p}=15$ seconds, as these conditions describe a typical 100-year return period at a typical site of interest. The reference point can be viewed as a possible position of a wind turbine, and we are interested in estimating the short-term exceedance probability of the quantity of interest for this location, see Fig. \ref{kdv_draw}. Therefore, the quantity of interest $g(\theta)$ is here the maximum crest elevation $\eta_{\rm max}$ within $T=600$s at the reference point $x^*$,
\begin{equation}
    \eta_{\rm max} = \max\{ \eta(x^*,t), 0\leq t \leq T \}. 
\end{equation}
Initially, we generate $N_{\rm MC}=5\cdot10^4$ evaluations of KdV22 for iid $\theta$ drawn from $\pi_d$ to produce the reference probability density function for $\eta_{\rm max}$ as well the reference short-term exceedance probability for $\eta_{\rm max}$, see Fig.~\ref{glavno}. The predefined sea state for $\theta$ is selected with $H_S=6.8$m and $T_P=15$s. In Fig. \ref{wavemax}a we recognize a heavy-tailed distribution with $\mu_{\eta_{\rm max}} \approx 5.7$m. The dimensionality of the input parameter $\theta$ depends on the frequency resolution and the time duration. In this present study, for a 10-minute wave simulation, $\theta$ is defined within $\mathbb{R}^{302}$. It represents a complex and high-dimensional problem for which standard reliability and surrogate methods become impractical.

\subsection{Dimensionality Reduction}
We employ the dimensionality reduction \textbf{Algorithm \ref{al12}} for the quantity of interest $\eta_{\rm max}$ and estimate the corresponding matrix $\mathbf{G}_{\rm max}$ by using the relation
\[
M=\alpha_A k_A \log(d)
\]
proposed by Constantine~\cite[p. 35]{paul1} for the number of samples $M$ sufficient to estimate the covariance matrix well. We define heuristically the oversampling factor $\alpha_A = 2.45$ and are interested in the first $100$ eigenvalues, therefore $M=544$ for $d=302$. We experimented with different numbers of samples and found $M=544$ to provide a good balance between performance and accuracy.

\begin{figure}[ht]
    \centering
    \includegraphics[scale=0.2]{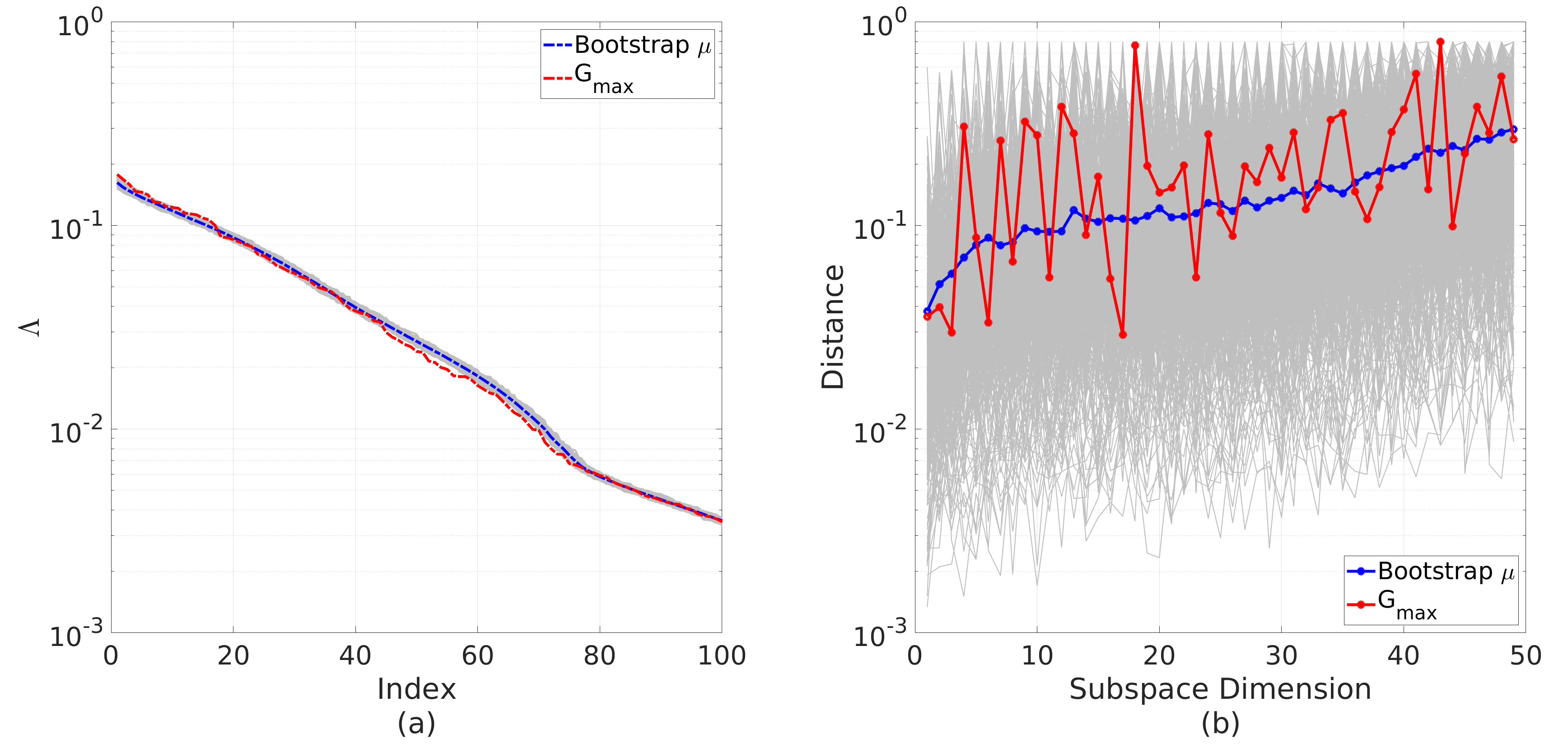}
    \caption{(a) The singular values $\boldsymbol{\Lambda}$ for the matrix $\mathbf{G}_{\rm max}$ from the active-subspace analysis with the 500 bootstrap replicates. (b) The estimated error in subspaces of dimension 1 to 49 with the 500 bootstrap replicates.}
    \label{wavemax}
\end{figure}

\begin{figure}[ht]
    \centering
    \includegraphics[scale=0.2]{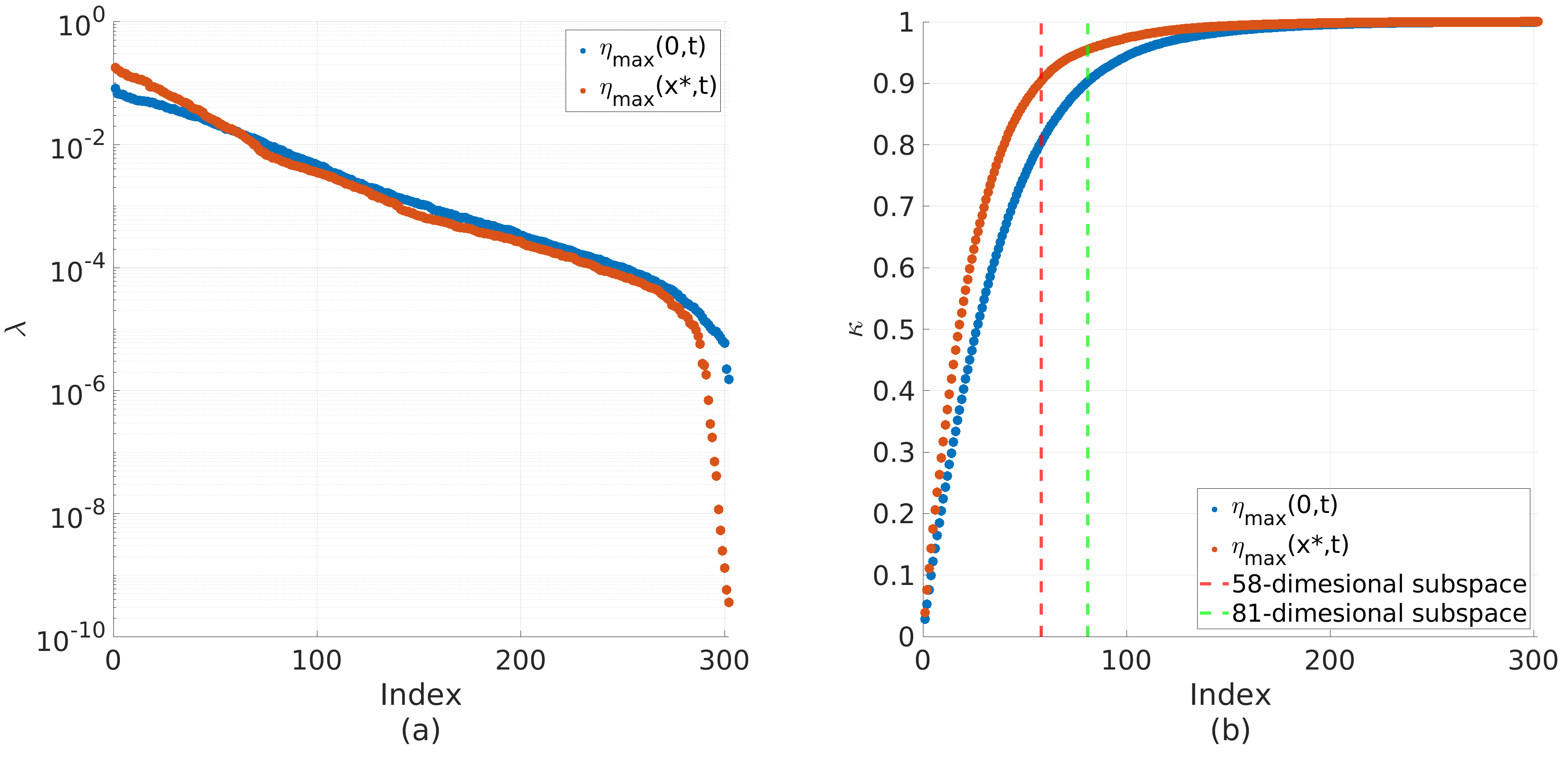}
    \caption{(a) The singular values $\boldsymbol{\Lambda}$ of the active-subspace analysis for the gradient matrices at the wave generation and the reference point $x^*$. (b) The ratio $\kappa$ between $\lambda_1 + ... + \lambda_r$ and $\lambda_1 + ... + \lambda_d$ with the green and red line as the $90\%$ threshold for $\eta_{\rm max}(0,t)$ and $\eta_{\rm max}(x^*,t)$.}
    \label{kap}
\end{figure}

Fig.~\ref{wavemax}a shows the singular values of the matrix $\mathbf{G}_{\rm max}$, and its corresponding bootstrap replicates. We notice relatively insignificant values, less than 0.2, for all singular values. Thus, the quantity of interest ($\eta_{\rm max}$) has low variability in each subspace direction. (Recall that a singular value expresses the expected variation of the square of the quantity of interest in the direction of its singular vector in input space.) Hence we do not need to sample significantly in these directions to have a good overall estimate of $\eta_{\rm max}$.
%This is probably because our model is weakly nonlinear.
%Similar behavior is found on the boundary before the weakly nonlinear propagation,
Fig. \ref{kap}a shows the effect of the weakly nonlinear propagation on the singular spectrum. Compared with the spectrum of the boundary condtion, the tail of the propagated spectrum shows an earlier decay at high frequencies and a seemingly more pronounced variation at low frequencies. This may indicate that a spectrum propagated by a fully nonlinear model, such as OceanWave3D~\cite{allan}, will feature a prominent spectral gap.
%Hence, following the relation between the singular values $\boldsymbol{\Lambda}$ at the wave generation and the reference point Fig. \ref{kap}a, we can assume that for a fully nonlinear model and a complex case with the slope, a spectral gap will be significant.
As the maximum frequency for the KdV22 wave model is $f_{\rm max}=0.2889$, the input parameters $\theta$ above $j=149$ in Eq.~\eqref{KC} are disregarded. This property is recognized as well in Fig. \ref{kap}a as the singular values above the index value of $289$ are insubstantial. The bootstrap replicates in Fig.~\ref{wavemax}a show the insignificant sample variation of the estimation.
%Henrik ima problem sa recenicom a spectral gap will be significant kao ne misli da je tacno sta li.  
As we cannot find a clear spectral gap in Fig. \ref{wavemax}a, we need to estimate the subspace errors using Eq.~\eqref{dist}. The upper bounds on the subspace errors, Fig. \ref{wavemax}b, suggest that the 17-dimensional subspace might be the optimal choice for $\eta_{\rm max}$. The bootstrap procedure in Figure~\ref{wavemax}b reveals a linear increase in the approximation error with increasing dimension. In view of Eq.~\eqref{dist}, this may be due to the overall flattening of the singular value spectrum with increasing index, that, due to $\lambda_r-\lambda_{r+1}$ approaching zero (on average) with increasing $r$.

\begin{figure}[ht]
    \centering
    \includegraphics[scale=0.26]{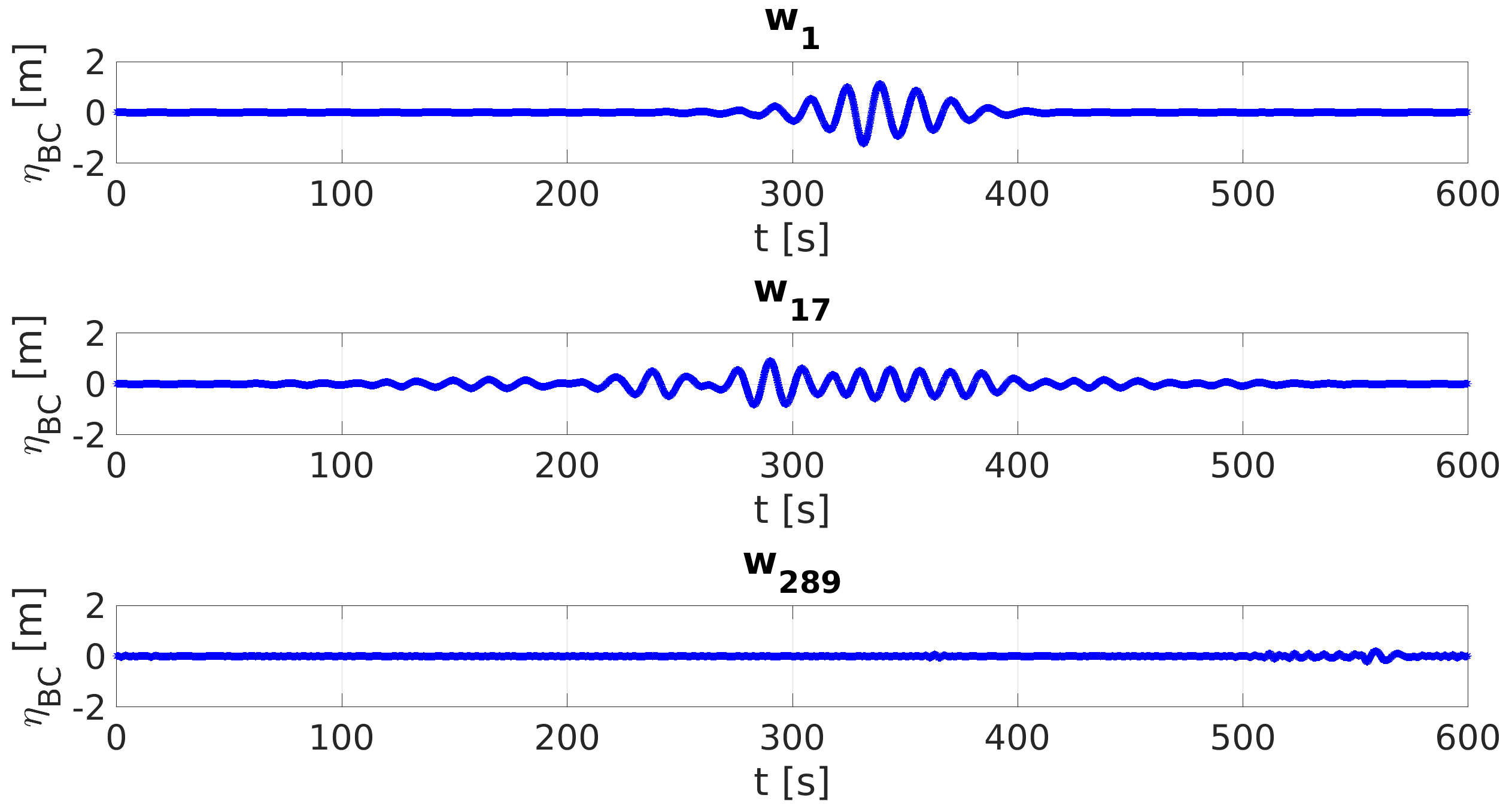}
    \caption{The surface elevations $\eta(x,t)$ at the boundary for the singular vectors $\mathbf{w}_{1}$, $\mathbf{w}_{17}$ and $\mathbf{w}_{289}$.}
    \label{w_waves}
\end{figure}
To additionally support our choice of the 17-dimensional subspace, we employ the coefficients of the singular vectors as the design parameters for the boundary condition in Eq.~\eqref{KC}. Figure \ref{w_waves} reveals the first singular vector $\mathbf{w}_1$ (the most active direction in the input space) to be a focused wave group, while the effect diminishes in singular vectors such as $\textbf{w}_{17}$ that correspond to smaller singular values (the less important directions). The singular vector $\mathbf{w}_{289}$ corresponds to the insignificant singular value $\lambda_{289}$, and it therefore represents insignificant free surface variations. It is well-known that extreme waves are associated with wave groups, cf. New Wave theory \cite{wavepack2,sapsis,wavepack1,wavepack3}. The ability of the active-subspace analysis to pick out initial conditions that produce a high degree of wave grouping at the structure thus confirms the relevance of the method. This way, we can construct active focused wave groups for future laboratory measurements. The singular vector $\mathbf{w}_{17}$ retains some of the localization, and it makes sense to keep it as an active direction. The spectrum above the index 17 is treated as measurement noise, for which Gaussian process regression is suitable \cite{paul2}.

The identification of the important directions can alternatively be based on a conservative approach~\cite{mirza} that uses the~\textit{total variation} of the singular values,
\begin{equation}\label{kappa}
    \kappa = \frac{\sum_{i=1}^r \lambda_i}{\sum_{i=1}^d \lambda_i}.
\end{equation}
The active-subspece dimension $r$ is then selected to preserve a certain percentage, say $90\%$, of the total variation, see Fig. \ref{kap}b. It is clear that the singular values with index above $150$ are negligible, and the variation $\kappa$ is preserved $100\%$. For practical reasons, we might select $90\%$ as our threshold, which would result in a 58-dimensional active subspace for the KdV22 model. In the following, we shall work both with a 17-dimensional and a 58-dimensional active subspace. It is interesting to note in Fig.~\ref{kap}b that the weakly nonlinear wave propagation decreases the dimension of the active subspace for the same level of total variation $\kappa$. We expect this effect to be even more prominent when using fully nonlinear models.

\begin{figure}[ht]
    \centering
    \includegraphics[scale=0.2]{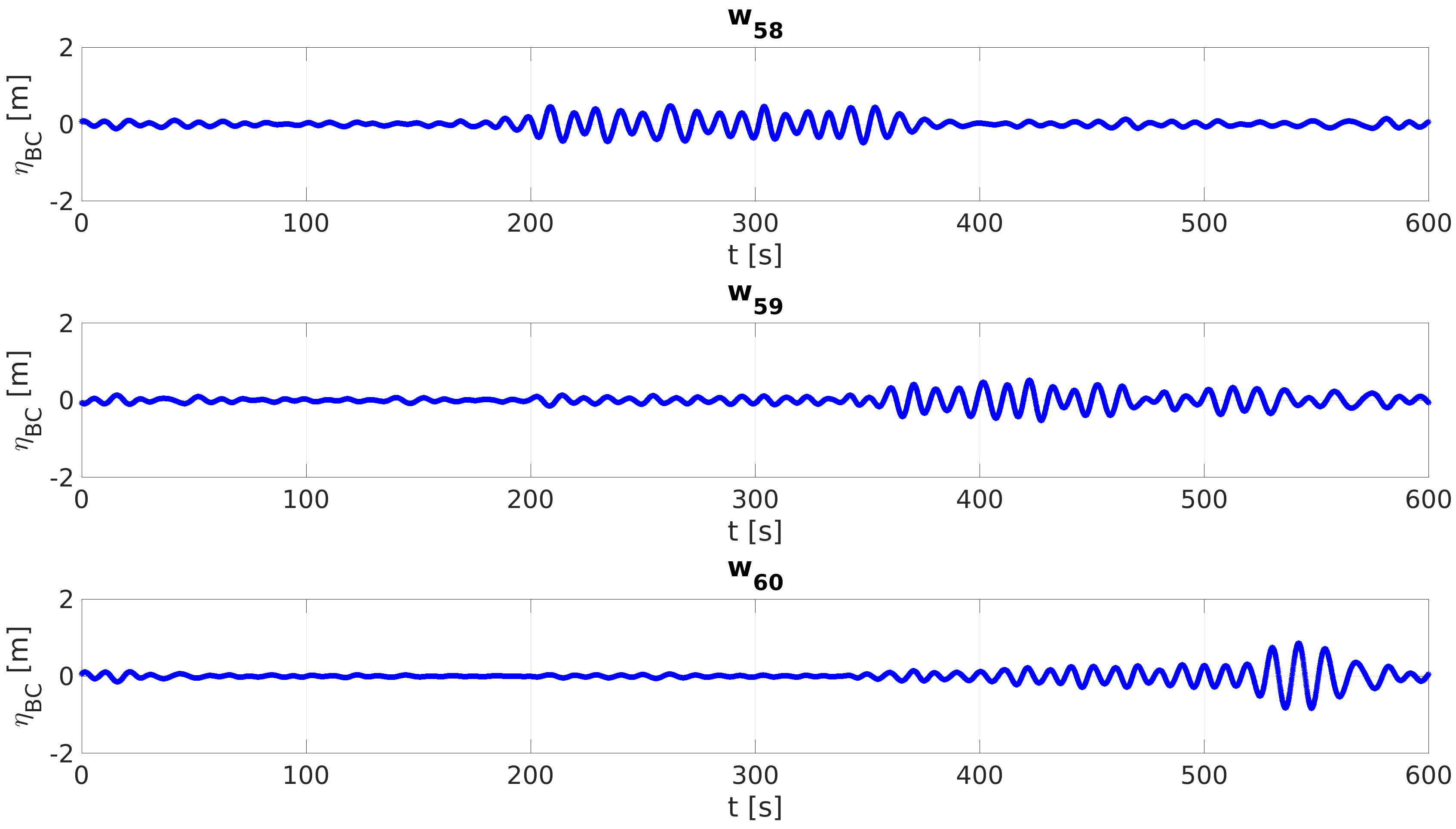}
    \caption{The initial surface elevations $\eta(x,t)$ at the boundary for the singular vector $\mathbf{w}_{58}$, $\mathbf{w}_{59}$ and $\mathbf{w}_{60}$}
    \label{w58}
\end{figure}

The directions of the singular vectors close to index 58 expose insignificant permutations of the initial surface elevation without clear wave groups, see Fig. \ref{w58}. Hence, the influence of these singular vectors on the overall result is insignificant as well, see Fig. \ref{kap}. For their singular values, we can expect that the Gaussian process architecture can easily control the error produced by neglecting the less important directions.

\subsection{Active-GP model}

The active-subspace analysis based on~\eqref{dist} and~\eqref{kappa} uses $17$-dimensional and $58$-dimensional active subspaces. We now construct the Gaussian process architecture on these low-dimensional subspaces, selecting the anisotropic squared exponential kernel which for the original high-dimensional space $\mathbb{R}^d$ is defined by
\begin{equation}
    K(|\theta_i-\theta_j|;\Theta) = \Theta_0\exp{\Bigg[-\frac{1}{2}\sum_{m=1}^d \frac{|\theta_{i,m}-\theta_{j,m}|^2}{\Theta_m}\Bigg]},
\end{equation}
where $\Theta=(\Theta_0,\dots,\Theta_d)$ are the hyperparameters. With the active low-dimensional projections $W_r^T\theta$, the kernel is now defined for $\mathbb{R}^r$ by
\begin{equation}
    K(|W_r^T\theta_i-W_r^T\theta_j|;\Theta) = \Theta_0\exp{\Bigg[-\frac{1}{2}\sum_{m=1}^r \frac{|(W_r^T\theta_i)_m-(W_r^T\theta_j)_m|^2}{\Theta_m}\Bigg]},
\end{equation}
which reduces the computation load since $r \ll d$.

The hyperparameters $\Theta$ are found using maximum likelihood estimation. As previously explained, typically a specific amount of variation of the quantity of interest is associated with each singular vector, with most variation occurring along the first singular vector. Thus, an anisotropic kernel is a natural choice. A squared exponential part is also a reasonable option due to the Gaussian property of ocean waves. The trend is based on the pure quadratic regression.

Our active-subspace analysis is based on $M=544$ evaluations $\eta_{{\rm max}, i} = g(\theta_i)$, and their input parameters $\theta_i$ are split randomly into the mutually disjoint training set and test set. The size of the training set depends on the active-subspace dimension. The rest of the samples are used as test cases. As we mentioned previously, finding the number of samples to be used for active-GP regression is a well-known problem. We used $N^{17}_{GP}=100$ for the 17-dimensional subspace and $N_{GP}^{58}=200$ for the conservative approach. We do not claim that this choice is the most efficient and accurate one. 

\begin{figure}[ht]
    \centering
    \includegraphics[scale=0.2]{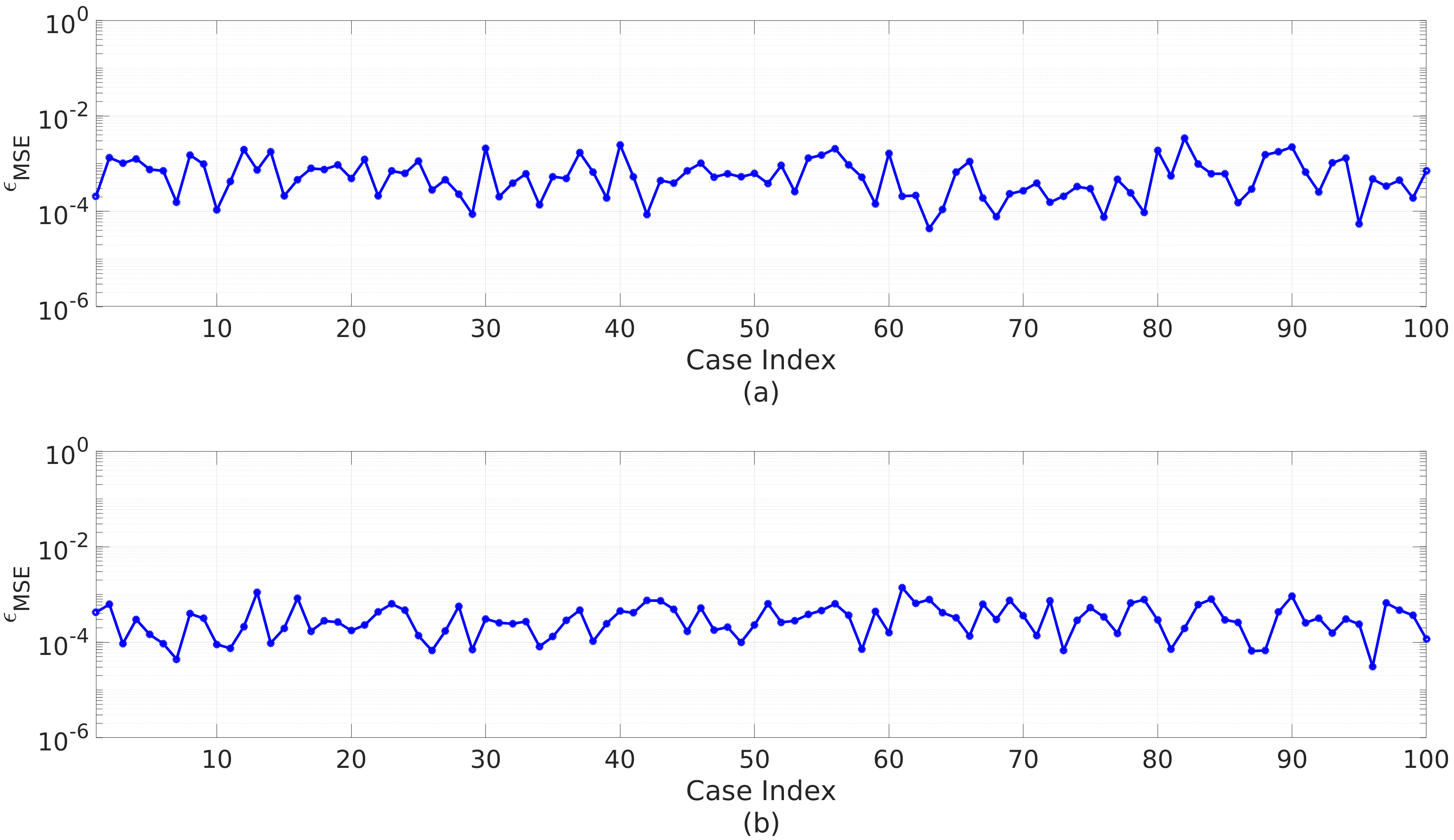}
    \caption{The mean-squared error estimations $\epsilon_{MSE}$ for the cross-validation tests of (a) 17-dimensional active-GP model and (b) 58-dimensional active-GP model.}
    \label{asa_mse}
\end{figure}

\begin{figure}[ht]
    \centering
    \includegraphics[scale=0.2]{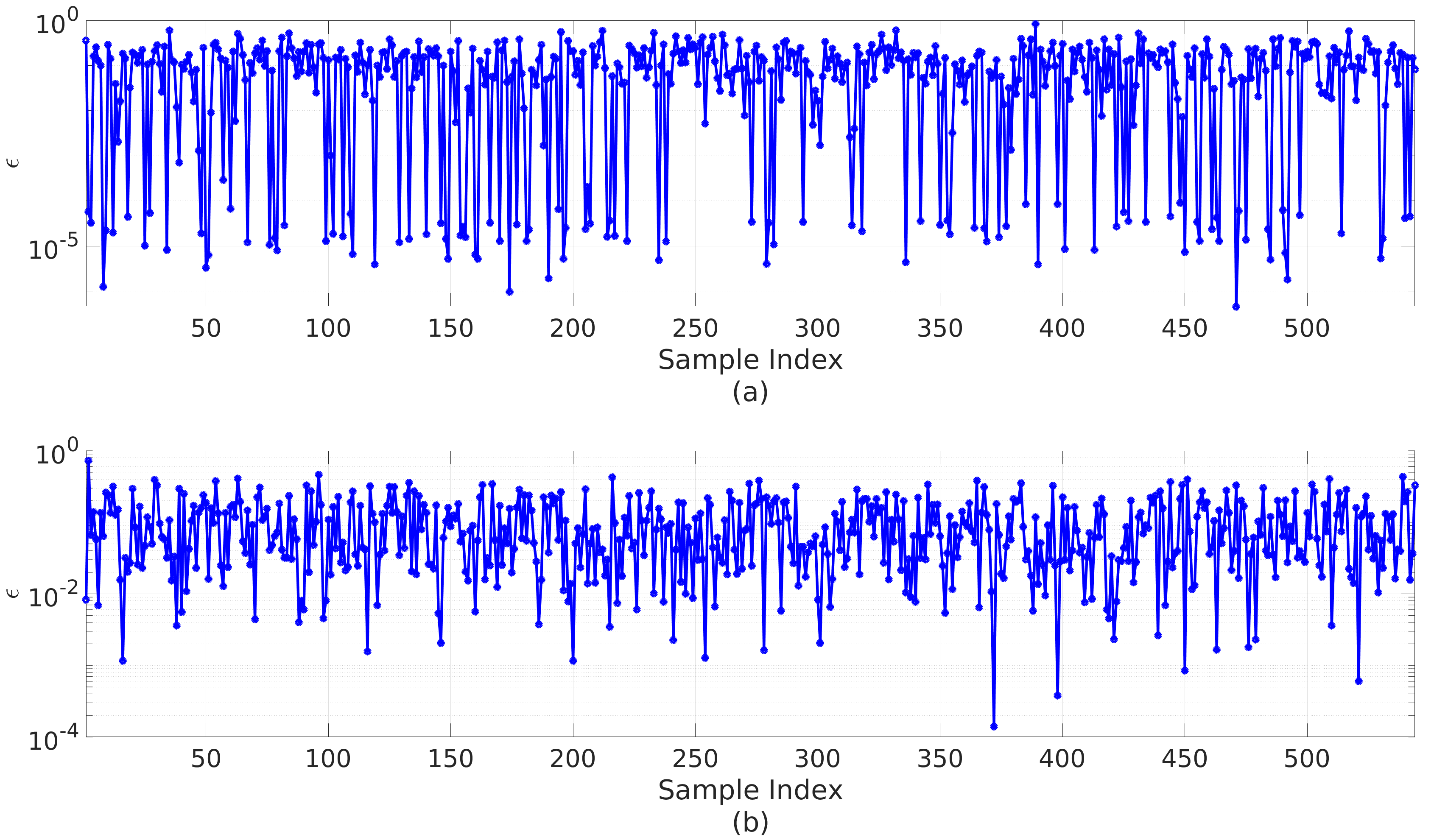}
    \caption{The relative error estimations $\epsilon$ for (a) 17-dimensional active-GP model and (b) 58-dimensional active-GP model.}
    \label{asa_rez}
\end{figure}

Based on the cross-validation procedure, we draw randomly $100$ distinct $N_{\rm GP}$-combinations of design points $\theta_i$ from the $M$ initial observations, and we also record the corresponding evaluations $\eta_{{\rm max},i}$. For each drawn combination, we train an active-GP model and estimate the mean-squared error (MSE) for the short-term exceedance probability based on the test data, see Fig. \ref{asa_mse}. We select the optimal design set that achieves a minimal MSE. The corresponding active-GP model is kept and used to evaluate all $M$ samples used in the active-subspace analysis, see Fig. \ref{asa_rez}. This figure shows the relative error in the predictions against the true evaluations for the $M$ samples. As we need to increase the sample set to $N^{58}_{GP}=200$ for the $58$-dimensional subspace, the performance of the active-GP models is not directly comparable. However, we can discuss the overall performance. The $17$-dimensional active-GP model based on the optimal cross-validation design set attains the relative error of $\approx13\%$  on average, which is for $\approx 15\%$ less than the relative error on average for the $58$-dimensional active-GP model. The maximum peaks of the relative error for these GP models are at $\approx 0.85$ and $\approx 0.77$, respectively. By adding singular vectors up to $\mathbf{w}_{58}$, we build up a Gaussian process architecture that would require a bigger kernel matrix and more design points to describe $g(\theta)$ properly. This can easily give poor performance for relatively small numbers of samples. In light of the singular values in Fig. \ref{wavemax} and of the initial surface elevations for the less important directions in Fig. \ref{w_waves}, we know that our quantity of interest changes on average insignificantly in the directions spanned by $\textbf{w}_j$ with $j>17$. We can expect that the active-GP architecture will compensate for the errors in the less important directions, and that an active-subspace analysis based on Eq.~\eqref{dist} is sufficient for this work. 

\begin{figure}[ht]
    \centering
    \includegraphics[scale=0.2]{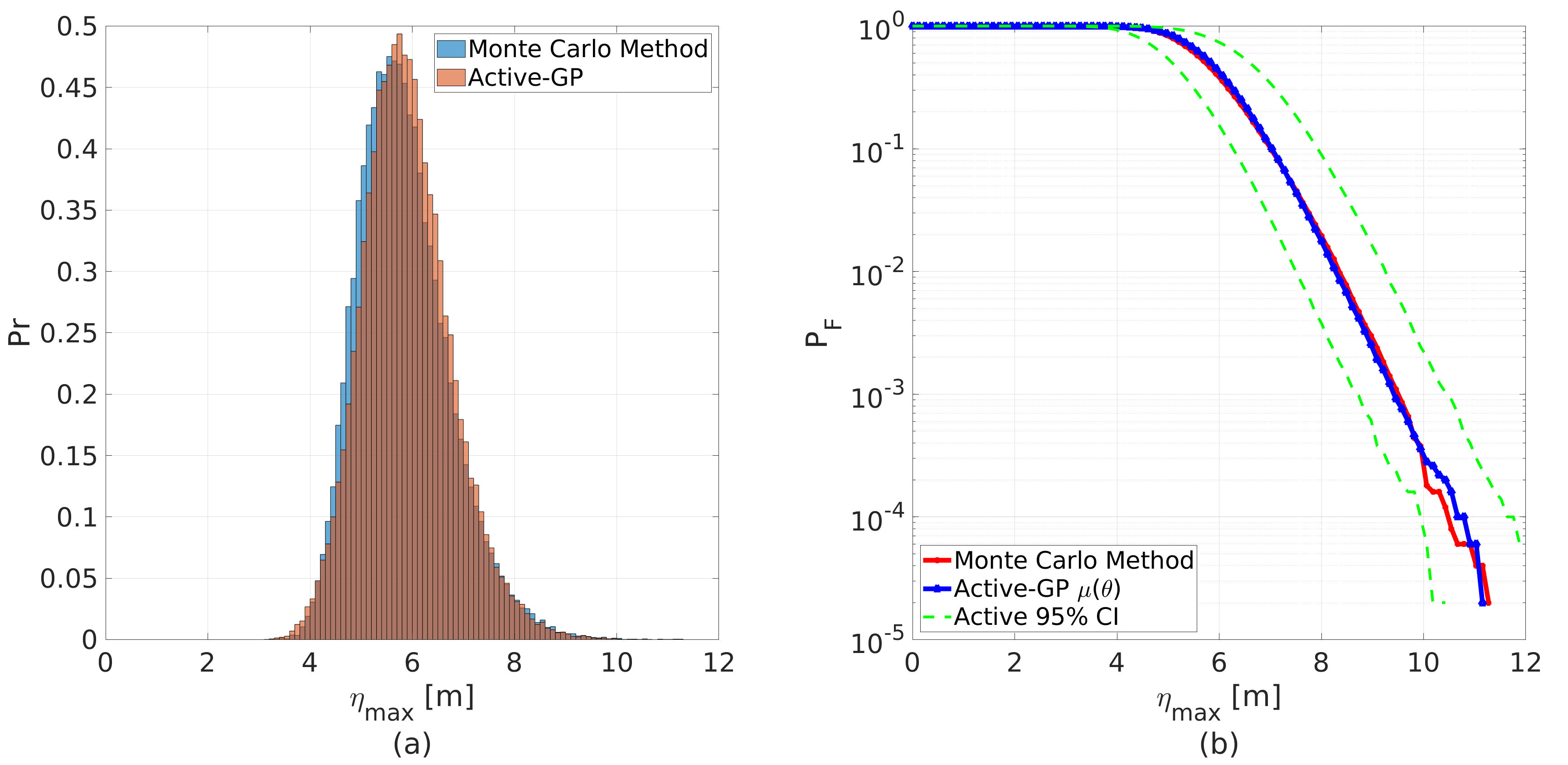}
    \caption{(a) The probability density function for $\eta_{\rm max}$. (b) The short-term exceedance probability of $\eta_{\rm max}$ for 10-minutes based on the active-GP model.}
    \label{glavno2}
\end{figure}

As we want to recreate the reference short-term exceedance probability, we evaluate the 17-dimensional active-GP model for $N=5\cdot 10^4$ and compare the performances with the simple Monte Carlo, see Fig. \ref{glavno2}. Figure \ref{glavno2}b demonstrates how well the active-GP model reproduces the performance of the simple Monte Carlo. The green lines are the $95\%$ confidence interval as a quality prediction measure because the Gaussian process method employs a distribution over the design points. This interval can be used in the sequential design to reduce the uncertainty in predictions \cite{bruno}. The active-GP model shows slight under-prediction around the exceedance order of $10^{-4}$ with the relative error of $6.3\%$ on average. The histograms, Fig.\ref{glavno2}a, are also almost identical with the $\ell^2$-distance of 0.2. For wind turbines, the exceedance probability typically ranges between $10^{-3}$ and $10^{-4}$. Therefore, the maximum crest elevation $\eta_{\rm max}$ at $10^{-3}$ for the simple Monte Carlo is $\eta_{\rm max}^{MC} \approx 9.45$m. The active-GP model based on $N^{17}_{GP}=100$ points estimates the maximum crest elevation as $\eta_{\rm max}^{GP} \approx 9.45$m, which gives the relative error of $0.1\%$. For the exceedance level of $10^{-4}$, the simple MC estimates $\eta_{\rm max} \approx 10.5$m, while the active-GP model $\eta_{\rm max}^{GP} \approx 10.65$m with the relative error of $1.4\%$. These results are collected based on only $544$ evaluations, used to estimate the matrix $\mathbf{G}_{\rm max}$ and to design the active-GP model. This is a reduction in the number of evaluations of $99\%$ compared with simple Monte Carlo.

\subsection{A global sensitivity measure}\label{sa}
\begin{figure}[ht]
    \centering
    \includegraphics[scale=0.24]{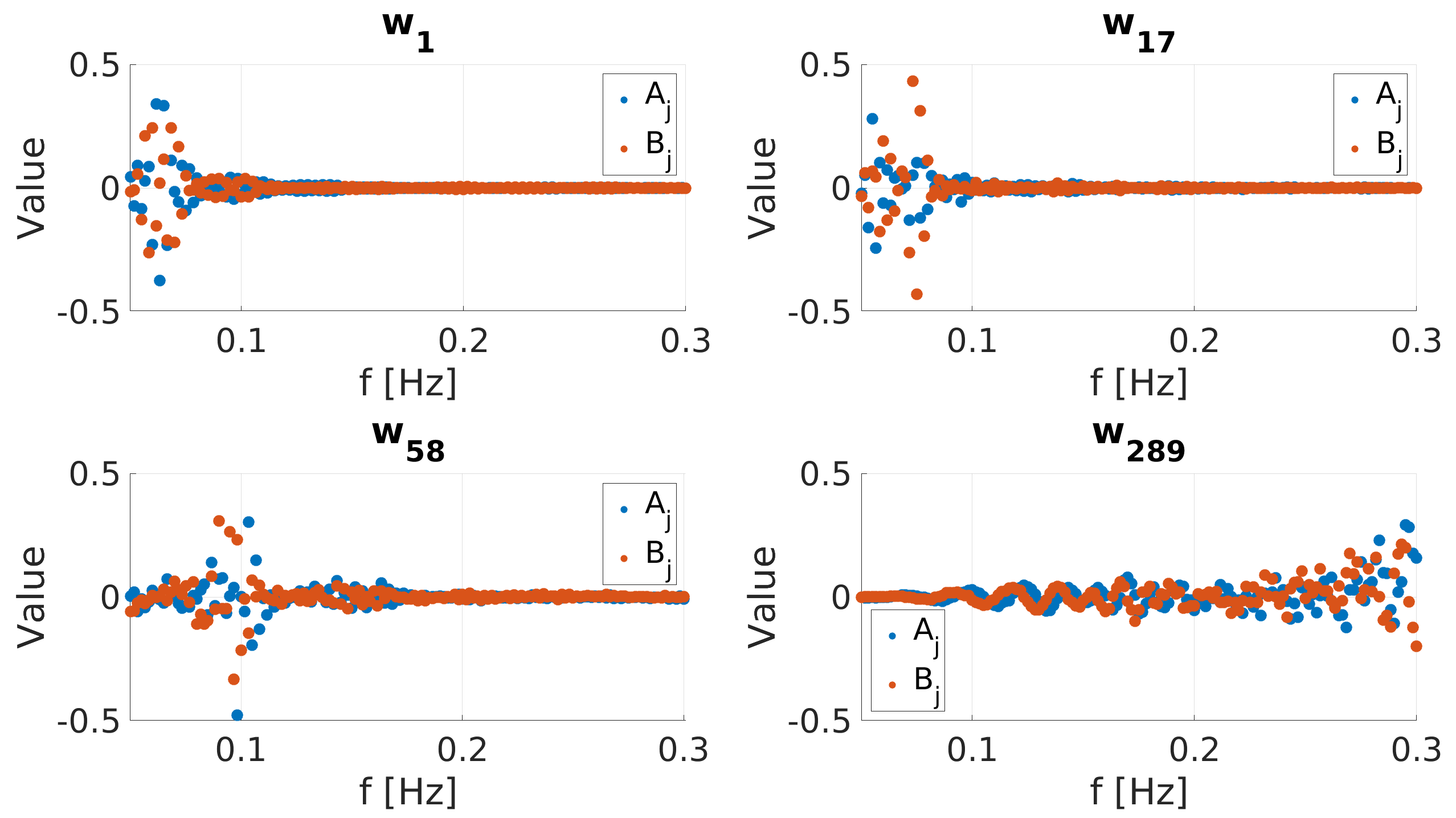}
    \caption{The components of singular vectors $\mathbf{w}_{1}$, $\mathbf{w}_{17}$, $\mathbf{w}_{58}$ and $\mathbf{w}_{289}$ for $A_j$ (blue) and $B_j$ (orange) as a function of the frequency $f$.}
    \label{w}
\end{figure}

Active-subspace analysis can also provide a sensitivity measure of the quantity of interest, $\eta_{\rm max}$, regarding the original input parameters $\theta$. In Fig. \ref{w} we plot the components of the singular vectors $\textbf{w}_j$ and their corresponding frequencies for $j=1$, 17, 58 and 289. We discover that the frequencies above $0.1$Hz are negligible for the singular vectors $\mathbf{w}_{1-17}$ that span the active subspace. This indicates that $66\%$ of the defined JONSWAP spectrum does not significantly affect the quantity of interest. The lower frequencies produce higher crests and deeper troughs, which will contribute most to the expectation value of $\eta_{\rm max}$. While moving in the directions of the less important vectors, e.g., $\mathbf{w}_{100}$ and $\mathbf{w}_{289}$, the higher frequencies (smaller waves) become more prominent, Fig. \ref{w}.

\begin{figure}[ht]
    \centering
    \includegraphics[scale=0.2]{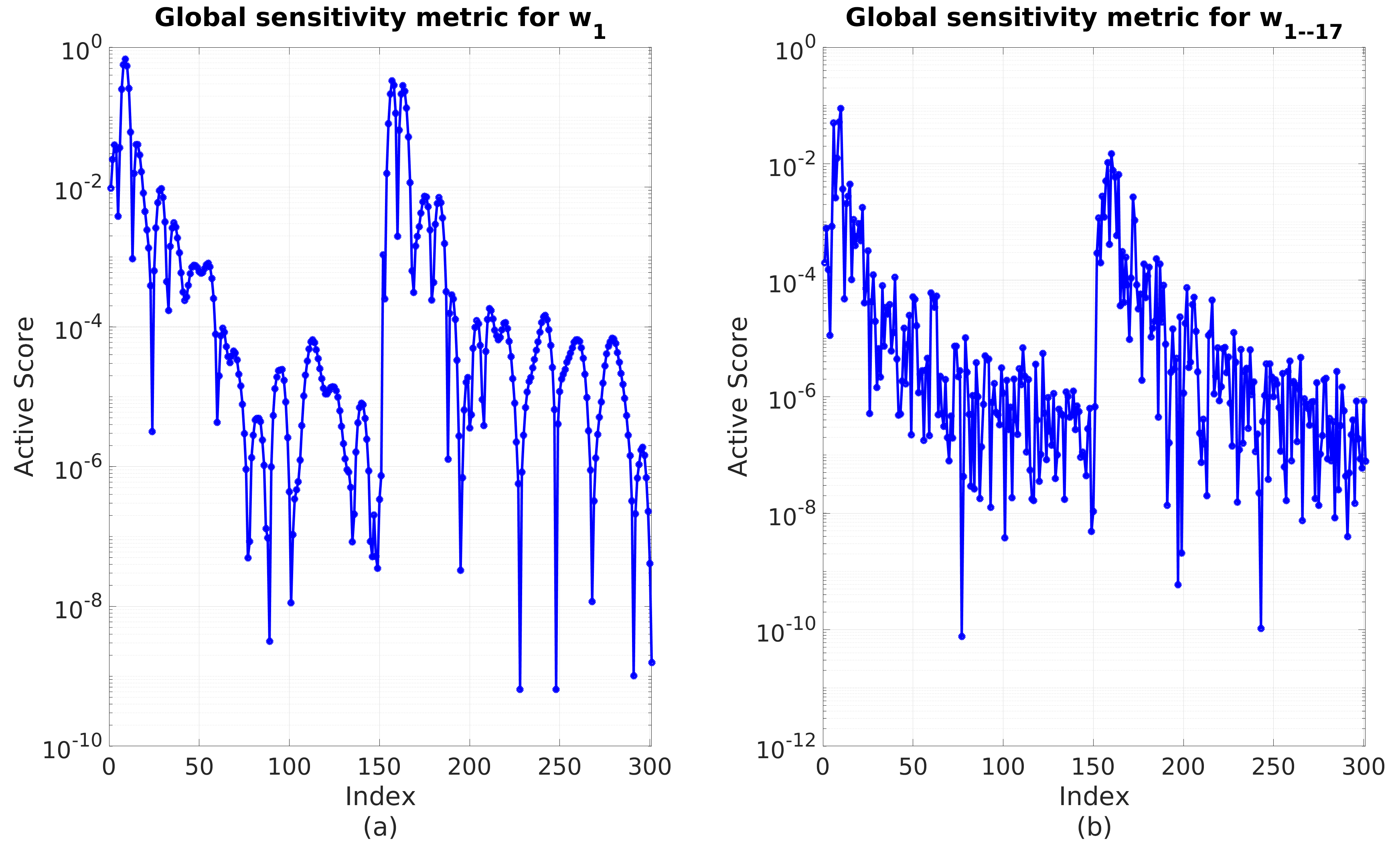}
    \caption{The activity scores for $A_j$ and $B_j$ for (a) $\mathbf{w}_1$ and (b) $\mathbf{w}_{1-17}$.}
    \label{score}
\end{figure}

We construct a global sensitivity metric, shown in Fig. \ref{score}, by multiplying the singular values $\lambda_j$, as the main indicator of the directional importance, with the squared components of the singular vectors. The so-called \textit{activity score} for the $j$'th component of the input $\theta$, or the $j$'th initial uncertainty parameter, is then defined by
\[
s_j = \sum_i^r \lambda_i\mathbf{w}_{i,j}^2,
\]
where $j\in \mathbb{R}^d$ \cite{score}. It is interesting to notice the second peak around the frequency $0.11$Hz for the input parameters with the index around $j=120$, see Fig. \ref{score}. A wave spectrum, such as the Jonswap spectrum, is typically a global sensitivity measure with respect to the initial uncertainties $A_j$ and $B_j$, $j=1,\dots,d/2$. We expect those $A_j$ and $B_j$ that correspond to the peak of a wave spectrum to be the most important input parameters, see Fig. \ref{score}. However, the second peak around $0.11$Hz can be related to the modification of the wave spectrum due to wave propagation. This behavior can be found in the offshore literature as well as \cite{turk}, which additionally proves the value of the active-subspace analysis.

\begin{figure}
    \centering
    \includegraphics[scale=0.2]{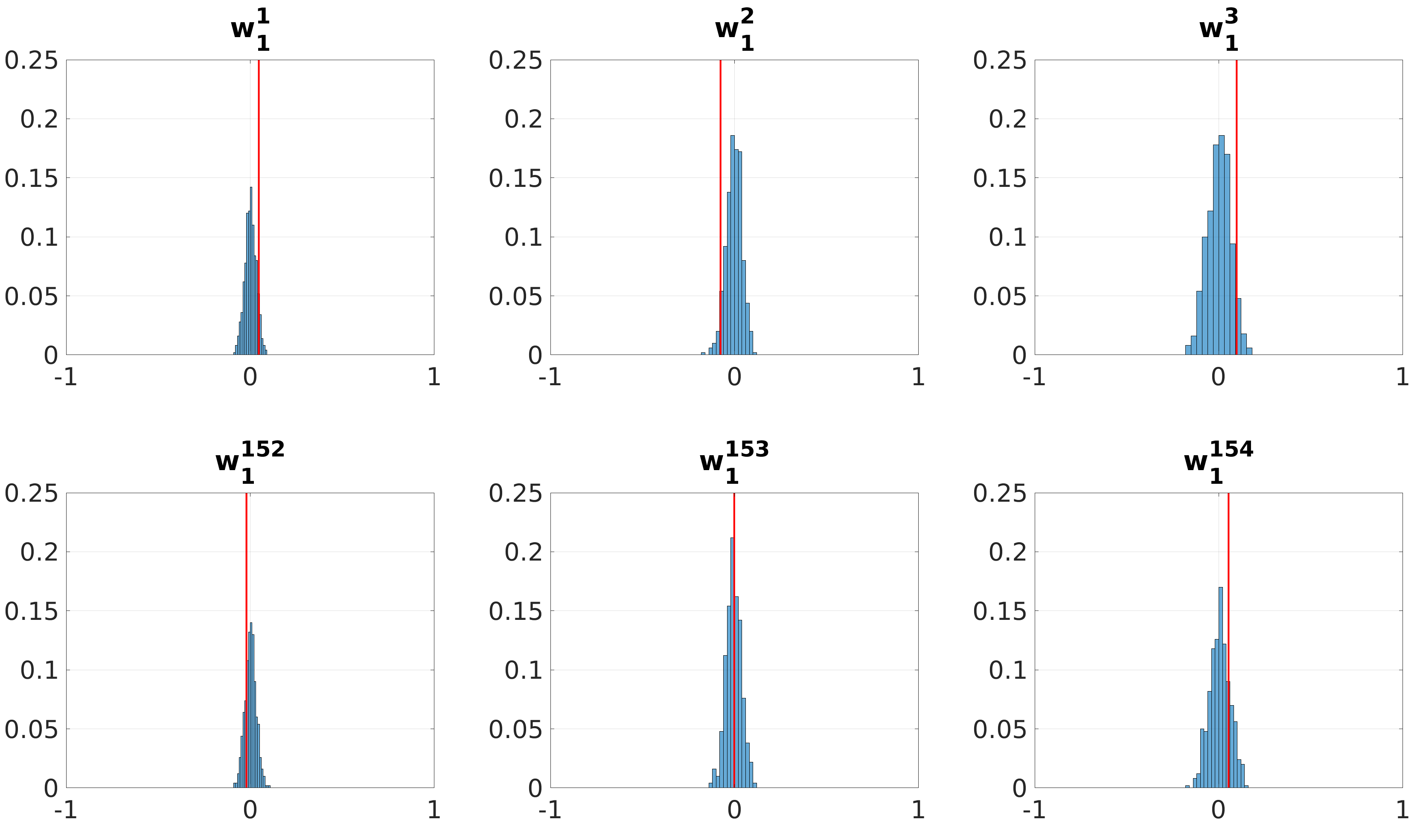}
    \caption{Bootstrap histograms of the components of the active subspace singular vector $\mathbf{w_1}$ for the maximum crest elevation $\eta_{\rm max}$.}
    \label{whist}
\end{figure}

To estimate the variability within, e.g., the components of $\mathbf{w}_1$, we employ the bootstrap approach with $500$ replicates for the covariance matrix $\mathbf{C}$ and \textbf{Line 5} of \textbf{Algorithm \ref{al12}}, see Fig. \ref{whist}. This cost is negligible because the bootstrap approach uses only the available model evaluations. The sharp peaks in the histogram around the expected value suggest confidence in the computed directions \cite{paul1}. The relatively wider histograms, see Fig. \ref{whist}, are due to the insufficient $M$.

\section{Conclusion}\label{section:end}
We apply a dimensionality reduction method called the active-subspace analysis (ASA) to a high-dimensional offshore problem. We model shallow-water waves using a simple but credible weakly nonlinear numerical model based on the Kortweg-de Vries equation (KdV22) with a high-dimensional initial Gaussian response. Our approach can be seen as an intermediate step toward a fully nonlinear model. For this high-dimensional complex problem, the standard offshore methods fail to provide accurate results or would have an infeasible convergence rate. The active-subspace analysis uses gradient evaluations to identify a low-dimensional subspace within the input space that is most significant in terms of the sensitivity of the output.

In contrast to Principal Component Analysis (PCA), the ASA reduces dimensionality while retaining information about the numerical model. However, estimating gradients is typically challenging and requires an adjoint solver for optimal efficiency. We perform our analysis using forward automatic differentiation despite the large required number of realizations. 

We apply the ASA to the maximum crest elevation at the reference point to reduce the uncertainty dimension at the wave generation within 10-minute wave propagation for a predefined sea state. The singular value decomposition (SVM) of the gradient evaluations reveals the slow spectral decay for the singular values without a clear spectral gap, which is crucial for accurate active subspace estimation. However, we can construct the low-dimensional active subspace based on the error bound, which exploits the relation between the true and estimated active subspace. Also, the active subspace exposes a focused wave group associated with extreme waves and loads. The global sensitivity of the ASA demonstrates the wave spectrum modification due to wave propagation. Based on the numerical evaluations used for SVM, we train efficiently Gaussian processes on the active subspace for different batches and select the Gaussian process with the lowest mean-squared error. Finally, by using the simple Monte Carlo method, the trained Gaussian process accurately estimates the short-term exceedance probability with the relative error of around $6\%$ on average. The reference short-term exceedance probability is obtained by $5\cdot10^4$ numerical evaluations, while the active-subspace analysis and Gaussian process regression use only $1\%$ of the required Monte Carlo evaluations to provide the comparable result efficiently. 

% \begin{acknowledgements}
\section*{Acknowledgements}
This research was funded by the DeRisk project of Innovation Fund Denmark, grant number 4106-00038B.
% \end{acknowledgements}

\bibliographystyle{unsrt}  
\bibliography{references}  %%% Remove comment to use the external .bib file (using bibtex).

\end{document}